\documentclass[%
reprint,
superscriptaddress,
showpacs,
 amsmath,amssymb,
 aps,
prl,
]{revtex4-1}

\newcommand\beq{\begin{equation}}
\newcommand\eeq{\end{equation}}

\usepackage{graphicx}
\usepackage{dcolumn}
\usepackage{bm}

\begin{document}


\title{\boldmath ${\cal PT}$ Metamaterials via Complex-Coordinate Transformation Optics}

\author{Giuseppe Castaldi}
\author{Silvio Savoia}
\author{Vincenzo Galdi}
\email{vgaldi@unisannio.it}
\affiliation{Waves Group, Department of Engineering, University of Sannio, I-82100 Benevento, Italy
}%

\author{Andrea Al\`u}
\affiliation{Department of Electrical and Computer Engineering, The University of Texas at Austin, Austin, TX 78712, USA}%

\author{Nader Engheta}
\affiliation{Department of Electrical and Systems Engineering, University of Pennsylvania, Philadelphia, PA 19104, USA}%

\date{\today}


\begin{abstract}
We extend  the transformation-optics paradigm to a complex spatial coordinate domain, in order to deal with electromagnetic metamaterials characterized by balanced loss and gain, giving special emphasis to {\em parity-time} ($\cal{PT}$) symmetric metamaterials. We apply this general theory to complex-source-point radiation and anisotropic transmission resonances, illustrating the capability and potentials of our approach in terms of systematic design, analytical modeling and physical insights into complex-coordinate wave-objects and resonant states.
\end{abstract}

\maketitle

Balanced loss-gain artificial materials have elicited a growing attention in optics and photonics, mostly inspired by the emerging {\em parity-time} ($\cal{PT}$) {\em symmetry} concept, which was originally introduced in connection with quantum physics \cite{Bender1} (see \cite{Bender4} for a comprehensive review). Against the traditional axioms in quantum mechanics, Bender and co-workers \cite{Bender4} proved that even non-Hermitian Hamiltonians may exhibit {\em entirely real} energy eigenspectra, as long as they commute with the combined $\cal{PT}$-operator and share the same eigenstates. This implies, as a necessary condition on the quantum potential, $V(-{\bf r})=V^*({\bf r})$, with ${\bf r}$ denoting a vector position and $^*$ complex conjugation. 

In view of the formal analogies between Schr\"{o}dinger and paraxial Helmholtz equations, the above concepts and conditions may be straightforwardly translated to scalar optics and photonics scenarios, with complex-valued refractive-index profiles $n(-{\bf r})=n^*({\bf r})$ playing the role of the quantum potential. Such symmetry condition cannot be found in natural materials, but it may be engineered within current metamaterial technology, with a judicious spatial modulation of optical gain and losses (either along or across the propagation direction). Besides providing convenient experimental testbeds for ${\cal PT}$-symmetry-induced quantum-field effects that are still a subject of debate, $\cal{PT}$-symmetric metamaterials constitute {\em per se} a very intriguing paradigm, as the complex interplay between losses and gain may give rise to a wealth of anomalous, and otherwise unattainable, light-matter interaction effects that extend far beyond the rather intuitive loss (over)compensation effects \cite{Shalaev1}. These include, for instance, double refraction \cite{Makris1}, power oscillations \cite{Makris1,Zheng1}, spontaneous $\cal{PT}$-symmetry breaking \cite{Guo1,Ruter}, beam switching \cite{Kivshar},
absorption-enhanced transmission \cite{Guo1}, effectively nonreciprocal propagation \cite{Ramezani1,Lin2,Ge1,Regensburger1,Mostafazadeh1,Longhi3,Longhi4}, spectral singularities \cite{Mostafazadeh2}, and coherent perfect absorption \cite{Longhi2,Chong2}, with perspective applications to new-generation optical components, switches, lasers, and absorbers.

In this Letter, we show that the {\em transformation optics} (TO) framework \cite{Leonhardt,Pendry} may be extended, via complex analytic continuation of the spatial coordinates, in order to deal with ${\cal PT}$-symmetric metamaterials. This extension brings along  
the powerful TO ``bag of tools,'' already applied successfully to a wide variety of field-manipulating metamaterials \cite{Chen}, in terms of systematic design, analytical modeling and valuable physical insights. Our approach
may be related to recent efforts in applying
the coordinate-transformation methods to quantum mechanics in order to generate classes of exactly-solvable ${\cal PT}$-symmetric potentials (see, e.g., \cite{Levai,Znojil} and references therein). 

For simplicity, we start considering an auxiliary vacuum space with Cartesian coordinates ${\bf r}'\equiv(x',y',z')$, where time-harmonic $[\exp(-i\omega t)]$ electric (${\bf J}'$) and magnetic (${\bf M}'$) sources radiate an electromagnetic (EM) field $({\bf E}',{\bf H}')$. We then consider a coordinate transformation 
\beq
{\bf r}'={\bf F}\left({\bf r}\right)
\label{eq:CT}
\eeq
into a new {\em curved-coordinate} space.
By relying on the covariance properties of Maxwell's equations, the TO framework \cite{Pendry} allows for a ``material'' interpretation of the field-manipulation effects induced by the coordinate transformation in (\ref{eq:CT}), in terms of a new set of sources (${\bf J}, {\bf M}$) and fields (${\bf E}, {\bf H}$) residing in a {\em flat} physical space ${\bf r}\equiv(x,y,z)$ filled up by an inhomogeneous, anisotropic ``transformation medium'' (characterized by relative permittivity and permeability tensors ${\underline {\underline \varepsilon}}$ and ${\underline {\underline \mu}}$, respectively) that are related to the original quantities as follows
\begin{subequations}
\begin{eqnarray}
\left\{
{\bf E},{\bf H}
\right\}
({\bf r}) &=& {\underline {\underline \Lambda}}^{T}({\bf r})
  \cdot \left\{
{\bf E}',{\bf H}'
\right\}
  \left[
{\bf F}\left({\bf r}\right)
  \right],\label{eq:field}\\
\left\{
{\bf J},{\bf M}
\right\}
({\bf r}) &=& 
\!\det\!\left[{\underline {\underline \Lambda}}
\left({\bf r}
\right)
\right]
{\underline {\underline \Lambda}}^{-1}\left({\bf r}
\right)
  \cdot \left\{
{\bf J}',{\bf M}'
\right\}
  \left[
{\bf F}\left({\bf r}\right)\right],\label{eq:source}\\
{\underline {\underline \varepsilon}}({\bf r})= {\underline {\underline \mu}}({\bf r})&=&
\!\det\!\left[{\underline {\underline \Lambda}}
\left({\bf r}
\right)
\right]
{\underline {\underline \Lambda}}^{-1}\left({\bf r}
\right)
  \cdot
{\underline {\underline \Lambda}}^{-T}\left({\bf r}
\right).
\label{eq:tensors}
\end{eqnarray}
\label{eq:transf}
\end{subequations}
\!\!\! In (\ref{eq:transf}), ${\underline {\underline \Lambda}}\equiv\partial(x',y',z')/\partial(x,y,z)$ indicates the Jacobian matrix of the transformation in (\ref{eq:CT}), while the symbol $\mbox{det}(\cdot)$ and the superscripts $^{-1}$ and $^{-T}$ denote the determinant, the inverse, and the inverse transpose, respectively.
From (\ref{eq:tensors}), it is evident that, in order for the resulting transformation medium to exhibit loss and/or gain, the coordinate transformation in (\ref{eq:CT}) must be {\em complex-valued}. Complex-coordinate extensions of TO have already been explored in connection with single-negative transformation media \cite{Castaldi} and field-amplitude control \cite{Popa1}. In the present investigation, although the framework can deal in principle with general (asymmetrical, unbalanced) loss-gain configurations, we focus on transformation media characterized by balanced loss and gain obeying the ${\cal PT}$-symmetry conditions.

First, it can be shown (see \cite{SuppMat} for details) that the necessary condition $n(-{\bf r})=n^*({\bf r})$, usually considered in the scalar case to achieve ${\cal PT}$ symmetry, can be generalized to our vector scenario as: ${\underline {\underline \varepsilon}}(-{\bf r})={\underline {\underline \varepsilon}}^*({\bf r})$ [or, equivalently, ${\underline {\underline \mu}}(-{\bf r})={\underline {\underline \mu}}^*({\bf r})$]. From (\ref{eq:tensors}), we observe that such conditions are automatically fulfilled if the coordinate transformation in (\ref{eq:CT}) is chosen so that
\beq
{\underline {\underline \Lambda}}({-\bf r})={\underline {\underline \Lambda}}^*({\bf r}).
\label{eq:PTS}
\eeq
Similar to the scalar case \cite{Bender4}, the condition in (\ref{eq:PTS}) is in general {\em not sufficient} to guarantee a real eigenspectrum. As a matter of fact, beyond a critical non-Herminicity threshold an abrupt phase transition (usually referred to as ``spontaneous $\cal{PT}$-symmetry breaking'') may occur for which the eigenspectrum becomes (partially or entirely) {\em complex} (see \cite{SuppMat} for details). It can be shown (see \cite{SuppMat} for details) that a TO-based ``${\cal PT}$ metamaterial'' generated via a
a {\em continuous} coordinate transformation (\ref{eq:CT}) [subject to (\ref{eq:PTS})] from the auxiliary vacuum space is inherently characterized by {\em full} ${\cal PT}$-symmetry, i.e., it does not undergo spontaneous symmetry breaking. Accordingly, this also implies that spontaneous $\cal{PT}$-symmetry breaking phenomena may be attained via suitably designed {\em discontinuous} coordinate transformations.

\begin{figure}
\begin{center}
\includegraphics [width=8cm]{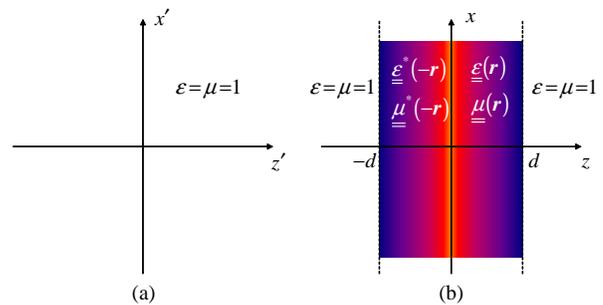}
\end{center}
\caption{(Color online) Problem schematic. (a) Auxiliary vacuum space. (b) Physical space with ${\cal PT}$-metamaterial slab, of thickness $2d$, associated with the coordinate transformations in (\ref{eq:CT1}).}
\label{Figure1}
\end{figure}

Our proposed TO framework allows the systematic synthesis [via (\ref{eq:tensors}) and (\ref{eq:PTS})] of ${\cal PT}$ metamaterials, as well as the analytical modeling of their EM response [via (\ref{eq:field}) and (\ref{eq:source})] and its physical interpretation in terms of analytically-continued complex-coordinate wave-objects residing in the auxiliary vacuum space. In what follows, for simplicity and without loss of generality, we focus on the two-dimensional (2-D) scenario illustrated in Fig. \ref{Figure1}, featuring a slab-type configuration associated with the coordinate transformation
\beq
x'=xu\left(z\right),~~~y'=y v\left(z\right),~~~z'=w\left(z\right),~~~|z|\le d,
\label{eq:CT1}
\eeq 
with the identity transformation implicitly assumed for $|z|>d$ (i.e., outside the slab region).

%

As a first example of application, we illustrate a TO-based ${\cal PT}$-metamaterial realization of the
{\em complex-source-point} (CSP) originally pioneered by Deschamps \cite{Deschamps} and Felsen \cite{Felsen} during the 1970s, and widely utilized to construct new classes of {\em exact} field solutions which convert point/line-source-excited
fields in a given environment into fields excited by Gaussian-beam-like wave objects in the same environment. In our example, we consider a unit-amplitude (V/m), $y$-directed magnetic line-source centered at $x'=0$ with a {\em purely imaginary} displacement along the $z'$ axis,
\beq
M_y'\left(x',z'\right)=\delta\left(x'\right)
\delta\left(z'-ib\right),~~b>0,
\label{eq:CSP}
\eeq
for which the $y$-directed radiated magnetic field  can be obtained via analytic continuation of the well-known 2-D Green's function \cite{FelsenBook}
\beq
H_y'\left(x',z'\right)=-\frac{\omega\varepsilon_0}{4}\mbox{H}_0^{(1)}\left(k_0{\tilde s'}\right),
\label{eq:CSP1}
\eeq
where $\mbox{H}_0^{(1)}$ denotes the zeroth-order Hankel function of first kind, $k_0=\omega\sqrt{\varepsilon_0\mu_0}=2\pi/\lambda_0$ is the vacuum wavenumber (with $\lambda_0$ denoting the corresponding wavelength), and ${\tilde s'}=\sqrt{x'^2+\left(z'-ib\right)^2},~~\mbox{Re}\left({\tilde s'}\right)\ge 0$,
represents a {\em complex distance}. This particular choice of branch-cut yields the so-called {\em source-type} solutions \cite{Felsen}, associated with an equivalent source distribution occupying the region $|x'|<b$ in the real $z'=0$ plane, and {\em discontinuous} across the same plane. This generates a wave-object that, within the paraxial regime (near the $z'$ axis), is well-approximated by a Gaussian beam with diffraction length $b$ propagating in the $z'>0$ halfspace, with only weak radiation for $z'<0$
(see \cite{SuppMat} for details).

%
\begin{figure*}
\begin{center}
\includegraphics [width=14cm]{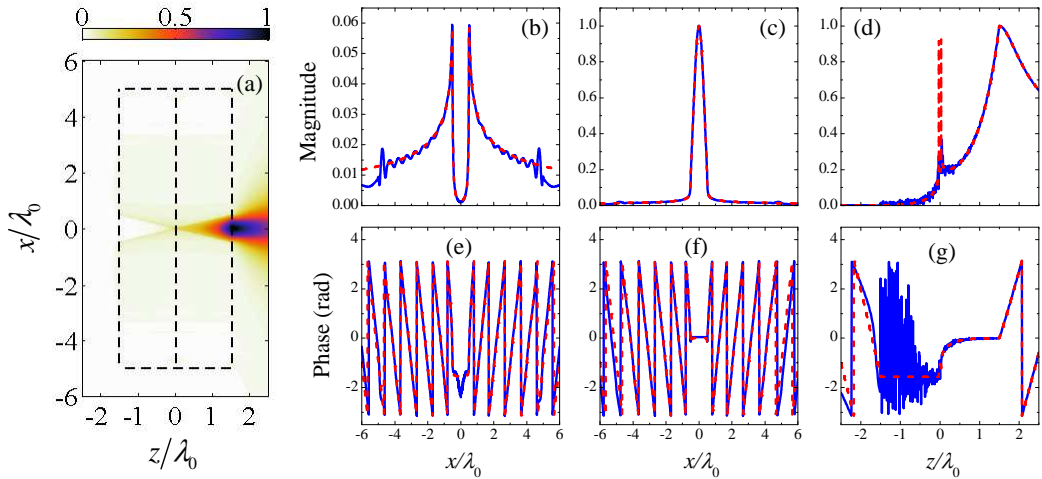}
\end{center}
\caption{(Color online) (a) Numerically-computed field magnitude ($|H_y|$) map  pertaining to a ${\cal PT}$ metamaterial slab of thickness $2d=3\lambda_0$,  transverse width $10\lambda_0$ (see dashed lines), and constitutive parameters as in (\ref{eq:MTMS1}) with $b=\lambda_0/2$ (i.e., $\varepsilon_{xx}=\mu_{yy}=\mp i/3$, $\varepsilon_{zz}=\pm 3i$, $z\gtrless 0$), excited by a magnetic line-source located at $x=z=0$. (b), (c), (d) Transverse and longitudinal magnitude cuts (blue-solid curves) at $z=\mp (d+\Delta)$ and $x=\Delta$, respectively, with the small displacement $\Delta=\lambda_0/100$ added so as to avoid branch-point and source-related singularities. (e), (f), (g) Corresponding phase profiles. Also shown (red-dashed curves) are the TO-based theoretical predictions [infinite slab; cf. (\ref{eq:field}) with (\ref{eq:CSP1})]. All field quantities are normalized with respect to $H_y(0,d)$.}
\label{Figure2}
\end{figure*}

Here, we show that this radiation phenomenon may be envisioned and realized in a physically appealing TO-based
${\cal PT}$-metamaterial slab excited by a line-source placed in a real-coordinate point. 
This is based on a simple coordinate transformation (see \cite{SuppMat} for details) 
\begin{subequations}
\begin{eqnarray}
\!\!\!\!\!u\left(z\right)&=&v\left(z\right)=1,\\
\!\!\!\!\!w\!\left(z\right)\!&=&\!ib\!\left(1\mp\frac{z}{d}\right)\!,~\mbox{Re}(z)\!\gtrless \!0,~|z|\!\le \!d,~\mbox{Im}(z)\!=\!0^+\!,
\label{eq:CT2}
\end{eqnarray}
\end{subequations}
which fulfills the conditions in (\ref{eq:PTS}) \cite{SuppMat}, and transforms [via (\ref{eq:source})] the original complex line-source (radiating in vacuum) into a conventional (real-coordinate) line-source embedded in a piecewise homogeneous ${\cal PT}$-metamaterial slab occupying the region $|z|\le d$. For the assumed transverse-magnetic (TM) polarization, the relevant nonzero constitutive-tensor components are purely imaginary,
\beq
\varepsilon_{xx}=\mu_{yy}=\mp \frac{ib}{d}, ~~\varepsilon_{zz}=\pm\frac{id}{b},~~z\gtrless 0, |z|\le d,
\label{eq:MTMS1}
\eeq
thereby representing a uniaxial zero-permittivity and zero-permeability metamaterial with balanced loss and gain. Note that, since the coordinate transformation in (\ref{eq:CT2}) is {\em continuous}, the metamaterial slab in (\ref{eq:MTMS1}) exhibits {\em full} ${\cal PT}$-symmetry.  
We stress that the chosen configuration serves only for illustrating, in the possibly simplest and more direct fashion, the general concept of metamaterial-induced wavefield ``complexification,'' which is a rather broad paradigm with a wealth of deep implications. Within this framework, no attempt was made to optimize the geometry and parameters so as to ensure the practical feasibility in a specific application scenario. Nonetheless, in \cite{SuppMat}, we do address some implementation and practical feasibility issues in connection with the ``exotic'' media described in (\ref{eq:MTMS1}). The final result is a metamaterial that, when excited by a line source, produces the exact beam field distribution originally described by Deschamps \cite{Deschamps} and Felsen \cite{Felsen} as a complex line source.
 
As an illustrative example, Fig. \ref{Figure2} shows the field map and representative cuts induced by a  line source embedded in a ${\cal PT}$-metamaterial slab as in (\ref{eq:MTMS1}), with relevant parameters given in the caption. For independent verification, we compare the theoretical TO-based predictions [cf. (\ref{eq:field}) with (\ref{eq:CSP1})] with results from full-wave numerical simulations (see \cite{SuppMat} for details) which also account for the finite extent (along $x$) of the slab. 
As it can be observed, the agreement is very good, apart for some small numerical oscillations of the phase in a region where the field amplitude is negligibly small. From a physical viewpoint, these results nicely illustrate how the ${\cal PT}$-metamaterial slab can physically implement the complex-source displacement, reproducing at its interfaces $z=\pm d$ the two branches of a discontinuous equivalent source distribution in the real $z'=0$ plane, continuously transitioning between them. 

%
\begin{figure*}
\begin{center}
\includegraphics [width=14cm]{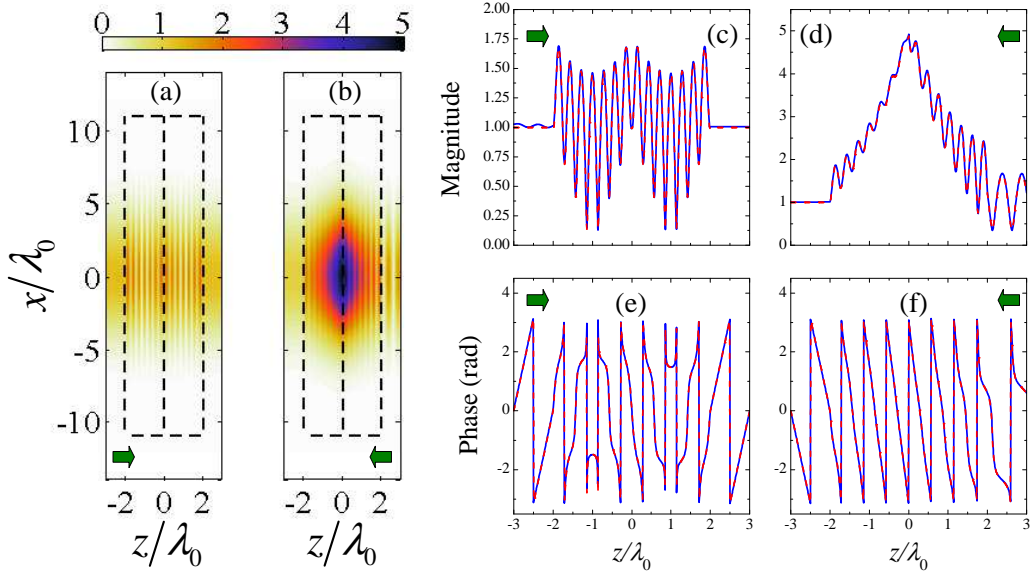}
\end{center}
\caption{(Color online) (a), (b) Numerically-computed field magnitude ($|H_y|$) maps  pertaining to a ${\cal PT}$-metamaterial slab of thickness $2d=4\lambda_0$,  transverse width $22\lambda_0$ (see dashed lines), and constitutive parameters as in (\ref{eq:MTMS1}) with $u_0=w_0=1.758-0.102 i$ and $v_0=3.079-0.359 i$ (i.e., $\varepsilon_{xx}=\varepsilon_{zz}=3.079\mp 0.359 i$, $z\gtrless 0$, and $\mu_{yy}=1$), excited by a unit-amplitude Gaussian beam with minimum waist of $4\lambda_0$ normally-incident from left and right, respectively (as schematically indicated by the thick arrows). (c), (d) Magnitude cuts (blue-solid curves) along the beam axis ($x=0$), for incidence from left and right, respectively. (e), (f) Corresponding phase profiles. Also shown (red-dashed curves) are the TO-based theoretical predictions (infinite slab and plane-wave incidence) \cite{SuppMat}.}
\label{Figure3}
\end{figure*}

As a second illustrative example of the relevance of this complex-coordinate TO paradigm, we present a metamaterial slab supporting
``anisotropic transmission resonances'' (ATRs) for which, at a single frequency, zero-reflection (with no phase accumulation) occurs only for incidence from one side of the structure and not from the other. Although these exotic propagation effects may be attained in more general non-Hermitian media (see, e.g., the Darboux-transformation approach \cite{Longhi5}), they have recently elicited a significant attention  within the ${\cal PT}$-symmetry context \cite{Ramezani1,Lin2,Ge1,Regensburger1,Mostafazadeh1,Longhi3,Longhi4}, and have been associated with symmetry-breaking phenomena. Accordingly, we consider a coordinate transformation as in (\ref{eq:CT1}) [subject to (\ref{eq:PTS})] with a generally {\em discontinuous} character
\begin{eqnarray}
u\left(z\right)&=&u_0^*,~v\left(z\right)=v_0^*,~w\left(z\right)=w_0^*z~,~-d\le z< 0,\nonumber\\
u\left(z\right)&=&u_0,~v\left(z\right)=v_0,~w\left(z\right)=w_0z,~~0< z\le d,
\label{eq:CT3}
\end{eqnarray}
which yields a ${\cal PT}$-metamaterial slab as in Fig. \ref{Figure1}.
We consider a plane-wave illumination along the $z$-axis, with $y$-directed magnetic field
\beq
H_{iy}^{(l,r)}(z)=\exp\left(\pm i k_0 z\right),
\eeq
with the $l,r$ superscripts denoting incidence from left and right (i.e., $+$ and $-$ sign in the exponential), respectively. It is possible to analytically calculate the slab response by following the approach in \cite{Castaldi} (see \cite{SuppMat} for details). In particular, letting $H^{(l,r)}_{ry}$ and $H^{(l,r)}_{ty}$ the corresponding reflected and transmitted fields, respectively, we focus on the reflection ($R_{l,r}$) and transmission ($T_{l,r}$) coefficients for incidence from left and right,
\begin{subequations}
\begin{eqnarray}
R_l&\equiv& \frac{H_{ry}^{(l)}(-d)}{H^{(l)}_{iy}(-d)},~~
R_r\equiv\frac{H_{ry}^{(r)}(d)}{H^{(r)}_{iy}(d)},\\
T_l&\equiv& \frac{H_y^{(l)}(d)}{H^{(l)}_{iy}(-d)}=T_r\equiv\frac{H_y^{(r)}(-d)}{H^{(r)}_{iy}(d)}.
\end{eqnarray}
\end{subequations}
It can be shown (see \cite{SuppMat} for details) that a sufficient condition for zero-reflection and no-phase accumulation for incidence from left ($R_l=0, T_l=1$) is
\beq
v_0=u_0 \left[i\tan\left(
\frac{w_0 k_0d}{2}
\right)\right]^{\pm 1},
\label{eq:v0u0}
\eeq
where the $\pm 1$ exponent identifies two distinct solutions. Under these conditions, we obtain  for the reflection coefficient from right \cite{SuppMat}
\beq
R_r
=\mp 4 i\mbox{Im}\left[
\cos\left(
w_0 k_0d
\right)
\right].
\label{eq:Rr}
\eeq
The relationships above depend in a remarkably simple fashion on the transformation parameters $u_0$, $v_0$ and $w_0$ in (\ref{eq:CT3}) and the electrical thickness $k_0d$, and provide useful insights into the effect of the complex-coordinate mapping. We note from (\ref{eq:v0u0}) that solutions may exist only if $v_0\neq \pm u_0$, which confirms the anticipated {\em discontinuous} character of the transformation. As expected, for a {\em real} coordinate transformation, solutions may exist only for {\em complex} values of the electrical thickness $k_0d$, which are {\em not physical}. Similar to the previous complex-source example, our complex-coordinate TO approach allows straightforward mapping of these physically-inaccessible solutions onto excitable resonant modes in a physical (real-coordinate) space. More specifically, for a given (real-valued) electrical thickness $k_0d$ and a desired value of the reflection coefficient from right ($R_r$), we can obtain from (\ref{eq:Rr}) the required value of $w_0$ and subsequently, from (\ref{eq:v0u0}), the ratio $v_0/u_0$ which identifies a {\em continuous infinity} of possible solutions. In particular, it is evident from (\ref{eq:Rr}) that, in order to attain {\em unidirectional} zero-reflection (i.e., $R_r\neq 0$) the transformation parameter $w_0$ needs simply to have nonzero real and imaginary parts.

Also in this case, the arising transformation medium is a piecewise homogeneous ${\cal PT}$-metamaterial slab,
\begin{eqnarray}
\varepsilon_{xx}&=&\frac{v_0^*w_0^*}{u_0^*}, \varepsilon_{zz}=\frac{v_0^*u_0^*}{w_0^*},~~\mu_{yy}=\frac{u_0^*w_0^*}{v_0^*},~~-d\le z<0,\nonumber\\
\varepsilon_{xx}&=&\frac{v_0w_0}{u_0}, \varepsilon_{zz}=\frac{v_0u_0}{w_0},~~\mu_{yy}=\frac{u_0w_0}{v_0},~0<z\le d,
\label{eq:MTMS2}
\end{eqnarray}
which, for the assumed polarization, can be made effectively {\em nonmagnetic} (i.e., $\mu_{yy}=1$) by enforcing $u_0w_0=v_0$, and 
{\em isotropic} ($\varepsilon_{xx}=\varepsilon_{zz}$) by enforcing $u_0=\pm w_0$. This significantly simplifies the practical implementation, at the expense of reducing the degrees of freedom in the solutions, which are now related to the (countably infinite, for given $k_0d$) roots of the transcendental equations
\beq
w_0=\left[i\tan\left(
\frac{w_0k_0d}{2}
\right)\right]^{\pm1},
\eeq
without direct control of $R_r$ in (\ref{eq:Rr}). 

Figure \ref{Figure3} shows the numerically-computed \cite{SuppMat} field maps and relevant field cuts (blue-solid curves) for a Gaussian-beam  
exciting a nonmagnetic, isotropic, truncated ${\cal PT}$-metamaterial slab with moderate thickness ($2d=4\lambda_0$) and $\varepsilon_{xx}=\varepsilon_{zz}=3.079\mp 0.359 i$, $z\gtrless 0$. It is fairly evident the strong difference between the responses pertaining to incidence from left [nearly-zero reflection with no phase accumulation; cf. Figs. \ref{Figure3}(a), \ref{Figure3}(c) and \ref{Figure3}(e)] and right [sensible reflection ($|R_r|\approx 0.67$); cf. Figs. \ref{Figure3}(b), \ref{Figure3}(d) and \ref{Figure3}(f)]. Once again, these results are in very good agreement with the TO-based theoretical predictions (red-dashed curves) for infinite slab and plane-wave excitation. Also in this case, the parameters were mainly chosen for the sake of simplicity of illustration and visualization, and the resulting gain levels required turn out to be unrealistic within current technology. However, we did verify that, in line with similar scenarios in the literature \cite{Mostafazadeh1}, much more feasible levels of loss/gain may be traded-off for larger electrical-thickness values (see \cite{SuppMat} for details). Finally, a detailed study of the spontaneous symmetry-breaking phenomenon can be found in \cite{SuppMat}.

In conclusion, we have shown that complex-coordinate TO may be exploited for systematic generation, design and modeling of ${\cal PT}$ metamaterials for a variety of applications. As illustrated in our examples, the most attractive and interesting aspect of the proposed approach is the metamaterial-based transposition to an actual physical space of wave-objects and resonant states residing in complex-coordinate spaces.
Given the power and pervasiveness of analytic-continuation approaches in wave-physics, this is expected to open up a plethora of new intriguing venues for ${\cal PT}$ metamaterials. Accordingly, current and future studies are aimed at exploring more general transformation classes, as well as different geometries (e.g., cylindrical and spherical) and applications.

\acknowledgments{A.A. acknowledges the support of Dr. Arje Nachman under the AFOSR YIP Grant
No. FA9550-11-1-0009.}


\begin{thebibliography}{99}




\bibitem{Bender1}{C. M. Bender and S. Boettcher, \prl~{\bf 80}, 5243 (1998).}



\bibitem{Bender4}{C. M Bender, Rep. Prog. Phys. {\bf 70}, 947 (2007).}

\bibitem{Shalaev1}{S. Xiao,	V. P. Drachev, A. V. Kildishev,	X. Ni, U. K. Chettiar, H.-K. Yuan, and V. M. Shalaev, \nat~{\bf 466}, 735 (2010).}






\bibitem{Makris1}{K. G. Makris, R. El-Ganainy, D. N. Christodoulides, and Z. H. Musslimani, \prl~ {\bf 100}, 103904 (2008).}

\bibitem{Zheng1}{M. C. Zheng, D. N. Christodoulides, R. Fleischmann, and T. Kottos, \pra~{\bf 82}, 010103(R) (2010).}

\bibitem{Guo1}{A. Guo, G. J. Salamo, D. Duchesne, R. Morandotti, M. Volatier-Ravat, V. Aimez, G. A. Siviloglou, and D. N. Christodoulides, \prl~{\bf 103}, 093902 (2009).}

\bibitem{Ruter}{C. E. R\"uter, K. G. Makris, R. El-Ganainy, D. N. Christodoulides, M. Segev, and D. Kip, Nature Phys. {\bf 6}, 192 (2010).}

\bibitem{Kivshar}{A. A. Sukhorukov, Z. Y. Xu, and Y. S. Kivshar,
\pra~{\bf 82}, 043818 (2010).}

\bibitem{Ramezani1}{H. Ramezani, T. Kottos, R. El-Ganainy, and D. N. Christodoulides,
\pra~ {\bf 82}, 043803 (2010).}

\bibitem{Longhi3}{S. Longhi, Phys. Rev. A {\bf 81}, 022102 (2010).}

\bibitem{Lin2}{Z. Lin, H. Ramezani, T. Eichelkraut, T. Kottos, H. Cao, and D. N. Christodoulides,
\prl~ {\bf 106}, 213901 (2011).}

\bibitem{Longhi4}{S. Longhi, J. Phys. A: Math. Theor. {\bf 44}, 485302 (2011).}


\bibitem{Ge1}{Li Ge, Y. D. Chong and A. D. Stone, \pra~{\bf 85}, 023802 (2012).}

\bibitem{Regensburger1}{A. Regensburger, C. Bersch, M.-A. Miri, G. Onishchukov,	D. N. Christodoulides, and U. Peschel, \nat~{\bf 488}, 167 (2012).}

\bibitem{Mostafazadeh1}{A. Mostafazadeh, \pra~ {\bf 87}, 012103 (2013).}


\bibitem{Mostafazadeh2}{A. Mostafazadeh, \prl~{\bf 102}, 220402 (2009).}


\bibitem{Longhi2}{S. Longhi, \pra~{\bf 82}, 031801(R) (2010).}

\bibitem{Chong2}{Y. D. Chong, Li Ge, and A. D. Stone, \prl~{\bf 106}, 093902 (2011).}



\bibitem{Leonhardt}{U. Leonhardt, Science {\bf 312}, 1777 (2006).} 

\bibitem{Pendry}{J. B. Pendry, D. Schurig, and D. R. Smith, Science {\bf 312}, 1780 (2006).}

\bibitem{Chen}{H. Chen, C. T. Chan, and P. Sheng, Nature Materials {\bf 9}, 387 (2010).}


\bibitem{Levai}{G. L\'evai and M. Znojil, J. Phys. A: Math. Gen. {\bf 33}, 7165 (2000).} 

\bibitem{Znojil}{M. Znojil and G. L\'evai, Phys. Lett. A {\bf 271}, 327 (2000).}


\bibitem{Castaldi}{G. Castaldi, I. Gallina, V. Galdi, A. Al\`u, and N. Engheta, J. Opt. {\bf 13}, 024011 (2011).}

\bibitem{Popa1}{B.-I. Popa and S. A. Cummer, \pra~{\bf 84}, 063837 (2011).}

\bibitem{SuppMat}{Supplementary material, available online at \href{http://tinyurl.com/ckxbtc3}{http://tinyurl.com/ckxbtc3}.}


\bibitem{Deschamps}{G. A. Deschamps, Electron. Lett. {\bf 7}, 684 (1971).}

\bibitem{Felsen}{L. B. Felsen, Symp. Math. {\bf 18}, 39 (1976).} 

\bibitem{FelsenBook}{L. B. Felsen and N. Marcuvitz, {\em Radiation and Scattering of Waves}
(IEEE-Wiley, Piscataway, NJ, 1994).}


\bibitem{Longhi5}{S. Longhi, Phys. Rev. A {\bf 82}, 032111 (2010).}


\end{thebibliography}
\end{document}